\documentclass[a4paper,11pt]{article}
\usepackage{pos}
\renewcommand{\logo}{\relax}

\usepackage{orcidlink}
\usepackage[utf8]{inputenc} 
\usepackage[russian,english]{babel}
\usepackage{csquotes}
\usepackage{tabularray}
\usepackage{wrapfig}
\usepackage[font=footnotesize]{subcaption}

\usepackage{xcolor}
\definecolor{darkgreen}{rgb}{0,.5,0}
\definecolor{darkblue}{rgb}{0,0,.5}
\definecolor{darkred}{rgb}{.4,0,0}

\usepackage{tikz}
\usetikzlibrary{arrows.meta}
\usetikzlibrary{calc,positioning}
\usetikzlibrary{decorations.markings}
\usetikzlibrary{decorations.pathreplacing}
\tikzstyle{edge} = [black,line width=.3mm]
\tikzstyle{vertex}=[circle,minimum size=1.5mm, draw=black, fill=black, inner sep=0mm]

\tikzstyle{v} = [circle, draw=black, line width=.2pt, fill=black, inner sep=0pt, minimum size=1.5mm]
\tikzstyle{vb} = [circle, inner sep=0.1pt, fill=jred, minimum size=1mm]

\usepackage{pgfplots}
\pgfplotsset{
	compat=1.18,
	table/search path={plotdata},
}
\usepackage{bbm}
\usepackage{cleveref}

\newcommand{\hepp}{\mathcal{H}}
\newcommand{\abs}[1]{\left|#1\right|}

\DeclareMathOperator{\Aut}{Aut}
\newcommand{\firstsymanzik}{ \mathcal{U}}

\renewcommand{\d}{ \,\mathrm{d}}
\newcommand{\loopnumber}{\ell}

\newcommand{\period}{\mathcal{P}}

\newcommand{\mb}{\mathbbm}

\title{Statistics and asymptotics of subdivergence-free Feynman integrals in $\phi^4$ theory}

\author*[a]{Paul-Hermann Balduf}
\author[b]{Kimia Shaban}
\author[c]{Johannes Thürigen}

\affiliation[a]{\orcidlink{0000-0003-4475-3031} Mathematical Institute, University of Oxford, OX2 6GG, UK; and St. Peter's College Oxford }

\affiliation[b]{\orcidlink{0009-0009-4774-6671}   Department of Computer Science, University of Toronto, 	Toronto, Ontario, M5S 2E4, Canada }

\affiliation[c]{\orcidlink{0000-0002-6262-1430} Institute for Analysis and Numerics, University of Münster, Orléans-Ring 10, 48149 Münster, Germany}

\emailAdd{paul-hermann.balduf@maths.ox.ac.uk}
\emailAdd{paul@paulbalduf.com}
\emailAdd{kimia.shaban@uwaterloo.ca}
\emailAdd{johannes.thuerigen@uni-muenster.de}

	\abstract{Recent algorithmic improvements have made it possible to  evaluate subdivergence-free (=primitive=skeleton) Feynman integrals in $\phi^4$ theory numerically up to 18 loops. By now, all such integrals up to 13 loops and several hundred thousand at higher loop order have been computed. This data enables a statistical analysis of the typical behaviour of Feynman integrals at large loop order. We find that the average value   grows exponentially, but  the observed growth rate is accurately described by its leading asymptotics only upwards of 25 loops. This is in contrast with the $N$-dependence of the $O(N)$-symmetric $\phi^4$ theory, which is close to its large-order asymptotics already around 10 loops.
	
	Secondly, the distribution of integrals  has a largely continuous inner part but a few extreme outliers. This makes uniform random sampling inefficient. We find that the value of the integral is correlated with many features of the graph, which can be used for importance sampling. With a naive test implementation we obtained an approximately 1000-fold speedup compared with uniform sampling. This suggests that in future work, Feynman amplitudes at large loop order might be computed numerically with statistical methods, rather than through enumerating and evaluating every individual integral.

	The talk is based on the articles \cite{balduf_statistics_2023},  \cite{balduf_predicting_2024},   \cite{balduf_primitive_2024}. }

\FullConference{The European Physical Society Conference on High Energy Physics (EPS-HEP2025)\\
7-11 July 2025\\
Marseille, France\\}

\begin{document}
\maketitle

\section{Introduction}

Feynman integrals arise in the perturbative treatment of various interacting systems, such as quantum field theory and general relativity.   The order in the perturbation series is called \emph{loop order} and corresponds to the number of cycles, hence the size, of the Feynman graph. Much effort has been invested in the last decades to devise mathematical and numerical methods to compute individual Feynman integrals, and this continues to be an active area of research. 
In the present talk, we consider a slightly different question, namely we seek to understand how Feynman integrals behave at large loop order, and to what extent this knowledge can be exploited in computations.

Generically, a Feynman integral is a complicated function of masses and momenta, and its value depends on choices such as gauge or renormalization conditions. To simplify our analysis, in the present talk we only  consider   logarithmically divergent (=\emph{vertex-type}) graphs without subdivergences (  \emph{primitive}). These Feynman integrals depend on the overall kinematic scale as a simple logarithm, whose coefficient is called \emph{Feynman period} \cite{broadhurst_knots_1995}. The period is a finite, positive, real number, independent of kinematics or renormalization scheme; In dimensional regularization, it is the residue of the  $\frac 1 \epsilon$-pole of the unrenormalized integral. Thereby, the period is the contribution of the primitive graph to the beta function of the theory. Periods exist in every renormalizable quantum field theory, we restrict ourselves to massless $\phi^4$ theory in $D=4-2\epsilon$ dimensions. Let $G$ be a vertex-type primitive graph, its period is given by the parametric Feynman integral 
\begin{align}\label{def:period} 
	\period (G)&:=  \left( \prod_{e\in E_G} \int \limits_0^\infty  \d a_e\;   \right)   \; \delta \left( 1-\sum_{e=1}^{\abs{E_G}} a_e \right) \frac{1} {\firstsymanzik^{ 2}(\left \lbrace a_e \right \rbrace   )} \in \mb R.
\end{align}
Here, $\firstsymanzik$ is the first Symanzik polynomial \cite{bogner_feynman_2010}, a homogeneous polynomial in the Schwinger parameters $a_e$ associated to each edge $e\in E_G$. 
The name \emph{period} stems from the fact that this integral is a period in the sense of \cite{kontsevich_periods_2001}.
See \cite{schnetz_quantum_2010} for other equivalent integral representations.    More general Feynman integrals in $\phi^4$ theory differ from  \cref{def:period} in that their integrand has a numerator that depends on masses and momenta.

We will often consider an $O(N)$ symmetric version of $\phi^4$ theory, also known as vector model. In that case, the Feynman integral of every graph gets multiplied by an $O(N)$ symmetry factor, the circuit partition polynomial, while the  parametric integral \cref{def:period} itself is unchanged. This setup is reviewed e.g. in \cite{balduf_primitive_2024}.

\section{Graph count and period symmetries}\label{sec:graph_count}

An immediate question is how many distinct (non-isomorphic under automorphisms of internal edges and vertices) vertex-type Feynman graphs there are. The number of graphs multiplied by their symmetry factors $\frac{1}{\abs{\Aut(G)}}$ is well understood and can be computed analytically \cite{borinsky_renormalized_2017,balduf_primitive_2024}. In particular, this number grows factorially with the loop order. The actual count of graphs, without symmetry factors, can to date only be obtained by generating all graphs. \Cref{tab:graphcount} shows some numbers; we observe that there are more than $10^9$ vertex-type graphs already at $\loopnumber=15$ loops. To put these numbers into perspective, the last column of \cref{tab:graphcount} shows the number of planar graphs, which is much smaller: At high loop order, almost all graphs are non-planar. 
 
A priori, the period as given by \cref{def:period} could be a different number for every primitive graph.    A period \emph{symmetry} is a pair of non-isomorphic  graphs for which the numerical value of the period coincides, despite the Symanzik polynomial $\firstsymanzik$ and thus the integrand being distinct.

\begin{table}[htb]
	\centering
	\caption{Number of primitive decompletions (=subdivergence-free vertex-type graphs), completions,   and planar graphs versus loop order (compare \cite[Table~2]{balduf_statistics_2023}).}
	\begin{tblr}{vline{2-4}=solid,
			stretch=0,
			cell{1}{2}={font = \small},
			row{1}={valign=m},
		}
		\hline
		$\loopnumber$ &{Decompletions \\ (= primitive vertex-type graphs)}&    completions &  Planar decompletions \\
		\hline
		5 & 3 &2 &  2 \\
		6 & 10 &5 &  5 \\
		9 & 1688 & 227 & 235 \\
		12 & 1{,}081{,}529 & 81{,}305 & 14{,}596 \\
		15 & 1{,}393{,}614{,}379 &82{,}698{,}184 &  1{,}015{,}339 \\
		\hline
	\end{tblr}
	\label{tab:graphcount}
\end{table}

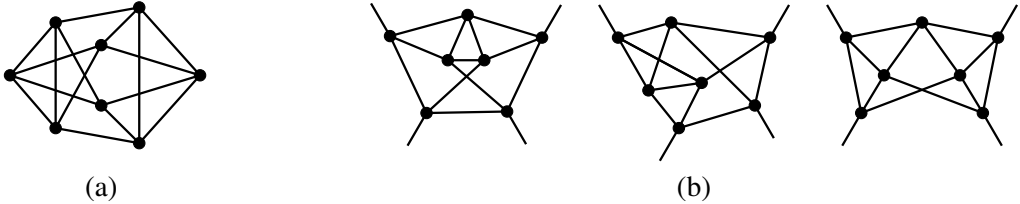
\begin{figure}[htb]
	\centering
	\begin{tikzpicture}[scale=1]
		
		\tikzset{vtx/.style={circle, fill=black, inner sep=1.2pt}}
		\coordinate (v0) at (0,  0);  
		\coordinate (v1) at (.6,  .7);  
		\coordinate (v2) at ( 1.2,  .4);  
		\coordinate (v3) at ( 1.7,  .9);  
		\coordinate (v4) at ( 1.7, -.9);  
		\coordinate (v5) at ( 1.2, -.4);  
		\coordinate (v6) at (.6, -.7);  
		\coordinate (v7) at ( 2.5,  0);  
		
		\draw[edge] (v0)--(v1) --(v3)--(v7) --(v4)--(v6) --(v0);
		\draw[edge] (v1)--(v6);
		\draw[edge] (v3)--(v4);
		\draw[edge] (v2)--(v0) -- (v5) --(v7)--(v2);
		\draw[edge] (v1) -- (v5) --(v4);
		\draw[edge] (v6) -- (v2) --(v3);
		
		\node[vertex] at (v0) {};
		\node[vertex] at (v1) {};
		\node[vertex] at (v2) {};
		\node[vertex] at (v3) {};
		\node[vertex] at (v4) {};
		\node[vertex] at (v5) {};
		\node[vertex] at (v6) {};
		\node[vertex] at (v7) {};
		
		\node at ($(v0)+(1.2,-1.5)$) {(a)};

		\coordinate (x0) at (5,  -.5);  
		\node at ($(x0)+(4,-1)$) {(b)};
		
		\coordinate (v0) at ($(x0)+ (0,1.02) $); 
		\coordinate (v1) at ($(x0)+ (.48,0) $); 
		\coordinate (v2) at ($(x0)+ (.76,.7) $);
		\coordinate (v3) at ($(x0)+ (1.02, 1.3) $);
		\coordinate (v4) at ($(x0)+ (1.24,0.70) $);
		\coordinate (v5) at ($(x0)+ (1.54,0.02) $);
		\coordinate (v6) at ($(x0)+ (2.00,1.00) $);
		\node[vertex] at (v0) {};
		\node[vertex] at (v1) {};
		\node[vertex] at (v2) {};
		\node[vertex] at (v3) {};
		\node[vertex] at (v4) {};
		\node[vertex] at (v5) {};
		\node[vertex] at (v6) {};
		
		\draw[edge](v0)--(v1)--(v4)--(v3)--(v2)--(v5)--(v6)--(v3)--(v0);
		\draw[edge] (v0)--(v2) -- (v4) -- (v6);
		\draw[edge](v1) -- (v5);
		\draw[edge] (v0) -- +(120:.5);
		\draw[edge] (v6) -- +(60:.5);
		\draw[edge] (v1) -- +(240:.5);
		\draw[edge] (v5) -- +(-60:.5);

		\coordinate (x0) at (8,  -.5);  
		
		\coordinate (v1) at ($(x0)+ (0,1.) $); 
		\coordinate (v2) at ($(x0)+ (.4,.3) $); 
		\coordinate (v3) at ($(x0)+ (.7,1.2) $);
		\coordinate (v4) at ($(x0)+ (2, 1) $);
		\coordinate (v5) at ($(x0)+ (1.1,0.4) $);
		\coordinate (v6) at ($(x0)+ (.8,-.2) $);
		\coordinate (v7) at ($(x0)+ (1.8,.1) $);
		\node[vertex] at (v1) {};
		\node[vertex] at (v2) {};
		\node[vertex] at (v3) {};
		\node[vertex] at (v4) {};
		\node[vertex] at (v5) {};
		\node[vertex] at (v6) {};
		\node[vertex] at (v7) {};
		
		\draw[edge] (v1) --(v3) --(v4)--(v7)--(v6)--(v2)--(v1);
		\draw[edge] (v1)--(v5)--(v4);
		\draw[edge](v3)--(v2)--(v5)--(v6);
		\draw[edge] (v3)--(v7);
		\draw[edge](v1) -- (v5);
		\draw[edge] (v1) -- +(120:.5);
		\draw[edge] (v4) -- +(60:.5);
		\draw[edge] (v6) -- +(240:.5);
		\draw[edge] (v7) -- +(-60:.5);

		\coordinate (x0) at (11,  -.5);  
		
		\coordinate (v1) at ($(x0)+ (0,1.) $); 
		\coordinate (v2) at ($(x0)+ (1,1.2) $); 
		\coordinate (v3) at ($(x0)+ (2,1) $);
		\coordinate (v4) at ($(x0)+ (.5, .5) $);
		\coordinate (v5) at ($(x0)+ (1.5,.5) $);
		\coordinate (v6) at ($(x0)+ (.2,0) $);
		\coordinate (v7) at ($(x0)+ (1.8,0) $);
		\node[vertex] at (v1) {};
		\node[vertex] at (v2) {};
		\node[vertex] at (v3) {};
		\node[vertex] at (v4) {};
		\node[vertex] at (v5) {};
		\node[vertex] at (v6) {};
		\node[vertex] at (v7) {};
		
		\draw[edge] (v1) --(v2)--(v3)--(v7)--(v5)--(v2)--(v4) --(v6)--(v1);
		\draw[edge] (v1) --(v4) --(v7);
		\draw[edge] (v3) -- (v5) -- (v6);
		\draw[edge] (v1) -- +(120:.5);
		\draw[edge] (v3) -- +(60:.5);
		\draw[edge] (v6) -- +(240:.5);
		\draw[edge] (v7) -- +(-60:.5);

	\end{tikzpicture}
	
	\caption{\textbf{(a)} A completion on eight vertices. It has eight decompletions, but all of them are isomorphic to either one of the three graphs shown in \textbf{(b)}. The decompletions have $\loopnumber=6$ loops. }
	\label{fig:completion-decompletion}
\end{figure}

The most prominent example of a symmetry is \emph{completion invariance}. A \emph{completion} is a subdivergence-free vacuum graph. By removing any one vertex, this graph becomes a vertex-type graph, called a \emph{decompletion}. Crucially, the same completion can give rise to multiple non-isomorphic decompletions, as shown in \cref{fig:completion-decompletion}. When $\loopnumber$ refers to the loop number of the decompletion, a completion has   $(\loopnumber+2)$ internal vertices. Thereby, completion invariance reduces the number of independent periods at loop order $\loopnumber$ by up to a factor of $(\loopnumber+2)$. To quantify this effect, we define $D_\text{rel}$ as the ratio between the number of non-isomorphic decompletions, and the total number of decompletions (which is $\loopnumber+2$). Hence, a value of $D_\text{rel}=1$ indicates that all decompletions are non-isomorphic. The graph in \cref{fig:completion-decompletion} has $D_\text{rel}=\frac 3 8$.

Completion invariance dramatically reduces the number of independent periods, as evidenced from the second column of \cref{tab:graphcount}. There are several further symmetries of the period \cite{schnetz_quantum_2010,hu_further_2022,schnetz_fivetwist_2025}, which together reduce the number of independent periods by another $\approx 20\%$, see precise counts in \cite{balduf_statistics_2023}. The remaining number of independent periods still exceeds one million at $\loopnumber \geq 14$, and one billion at $\loopnumber \geq 17$; this illustrates that despite all symmetries, it is practically impossible to compute the sum of periods at large loop order by computing each and every Feynman integral and summing them up. We demonstrate in \cref{sec:correlation} that clever random sampling is much more efficient.

It might appear that the property of having no subdivergences is very restrictive, but that is not the case. 
As demonstrated in \cite[Sec.~2.2]{balduf_statistics_2023}, the statistical properties of primitive completions are reasonably well described by random 4-regular simple graphs (i.e. graphs without double edges, but otherwise not restricted, not even necessarily connected) in the asymptotic limit   $\loopnumber\rightarrow \infty$.

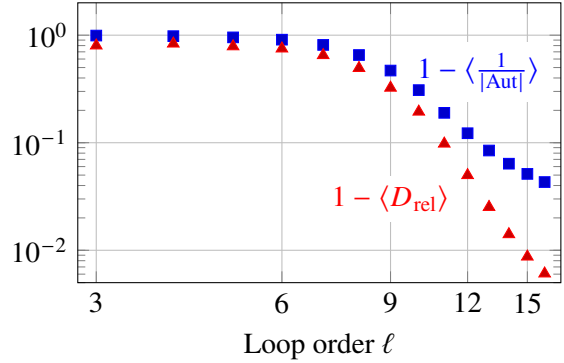
\begin{wrapfigure}[13]{R}{.5\linewidth}
	\centering
	\begin{tikzpicture}
		\begin{axis}[
			width=1.05\linewidth,
			height=0.7\linewidth,
			xmode=log,
			xlabel={Loop order $\loopnumber$},
			xmin=2.8, xmax=17, 
			xtick= {3,6,9,12,15},
			xticklabels={$3$,$6$,$9$,$12$,$15$},
			ymode=log,
			ymin=5e-3, ymax=2, 
			yticklabel style={
				anchor=east,
				/pgf/number format/precision=1,
			},
			yminorgrids=false,
			grid=major, 
			]

			\addplot+[only marks, mark size=2.0pt,mark=square*]
			t   table {plotdata/meanSymf.txt};
			\node[blue,fill=white] at (axis cs:12.5,.45){$1-\langle \frac 1 {\abs\Aut} \rangle$};

			\addplot+[only marks, mark size=2.5pt,mark=triangle*]
			table{plotdata/meanDrel.txt};
			\node[red,fill=white] at (axis cs:9,.03){$1-\langle D_{\rm rel}\rangle$};
			
		\end{axis}
	\end{tikzpicture}
	\vskip-.2cm
	\caption{Convergence of graph–symmetry indicators to the asymptotic regime, double logarithmic plot. The quantities quickly approach zero upwards of 10 loops.}
	\label{fig:aut_drel}
\end{wrapfigure}

\noindent
Many properties of such random graphs are known, e.g. \cite{wormald_asymptotic_1981,mckay_short_2004,mckay_automorphisms_1984,bollobas_diameter_1982,wormald_models_1999}.
In particular, in random graphs both the average $\langle D_\text{rel}\rangle$ and the average symmetry factor $ \langle \frac 1 {\abs{\Aut}}  \rangle$ approach unity when $\loopnumber \rightarrow \infty$. In \cref{fig:aut_drel}, we show the deviation of these quantities from unity, it rapidly diminishes above a certain threshold $\loopnumber \approx 9$ loops. One could thus be tempted to assume that Feynman integrals behave according to some large-order asymptotic law  for $\loopnumber\geq 10$ loops.   Conversely, we demonstrate in \cref{sec:asymptotics} that   the numerical value of periods behaves asymptotically only upwards of 25 loops.

\section{Numerical evaluation and Distribution}\label{sec:distribution}

\begin{figure}[htbp]
	\centering
	
	\includegraphics[width=.9\linewidth]{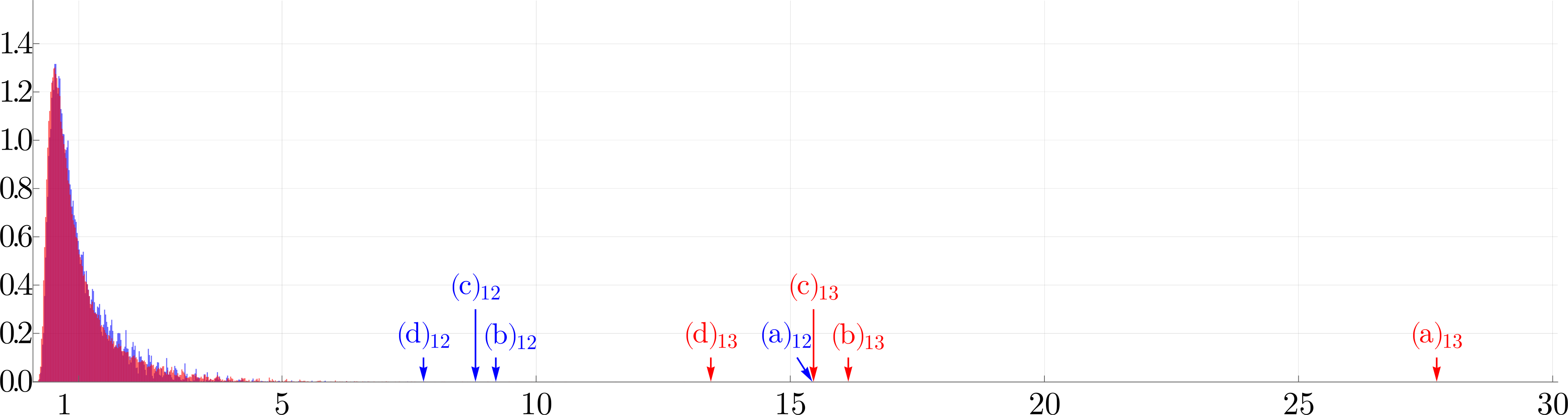}
	\caption{Distribution of periods at loop orders 12 (blue) and 13 (red), scaled to their respective mean. The histograms largely overlap, except for the outliers marked by arrows. Figure taken from \cite{balduf_statistics_2023}. }
	\label{fig:distribution}
\end{figure}

We computed more than 2 million periods numerically with the tropical integration algorithm of \cite{borinsky_tropical_2023a,borinsky_tropical_2023}; our data set includes all periods up to $\loopnumber \leq 13$ loops and partial random samples up to $\loopnumber \leq 18$. This data set is publicly available from \cite{balduf_feynman_2024} and the first author's website. 
The distribution of periods (not divided by their symmetry factor), normalized to unit mean, is shown in \cref{fig:distribution}. We observe that most of them lie relatively close to the mean, and this pattern does not depend on the loop order, see also the recent work  \cite{borinsky_feynman_2025} where a distribution function is conjectured. 

Conversely, there are few, but striking, outliers, marked by letters (a) to (d) in \cref{fig:distribution}. The relative magnitude of these outliers grows with the loop order. As these particular periods grow faster than the mean, they are not representative of the typical behaviour for the majority of periods at large loop order. Most notably, (a) is the so-called \emph{zig-zag-graph} \cite{broadhurst_knots_1995,brown_singlevalued_2015}, the only infinite family of periods in $\phi^4$ theory whose value is known analytically at all loop orders.

\section{Asymptotics of the period mean and the primitive $\beta$ function} \label{sec:asymptotics}

While the distribution of periods relative to their mean is relatively stable with increasing loop order, the mean itself is expected to grow exponentially, with a leading asymptotics given by \cite{mckane_perturbation_2019}  $\period \sim C \cdot \left(\frac 3 2\right)^\loopnumber \loopnumber^{\frac 5 2}$ with a known constant $C$. Our numerical data confirms an exponential growth, but still at $\loopnumber=18$, it is not possible to clearly infer the growth constants $\frac 2 3$ and $\frac 52$ from the data. This difficulty remains even if one additionally uses the pattern of $N$-dependence, as discussed in \cite[Sec.~4]{balduf_primitive_2024}. To visualize this situation, one considers the ratio $f_\loopnumber$ of subsequent loop orders, which is expected to be independent of $N$ up to the first subleading order \cite[Sec.~2.3]{balduf_primitive_2024}, 
\begin{align}\label{fL}
	f_\loopnumber :=  \textstyle{\left \langle \frac{\period}{\abs{\Aut}} \right \rangle_{\loopnumber+1}} \Big /  {\left \langle \frac{\period}{\abs{\Aut}} \right \rangle_\loopnumber} ~\displaystyle \sim \frac 3 2 + \frac{15}4 \frac 1 \loopnumber + \mathcal  O \left( \loopnumber^{-2}\right) \quad \text{as}\quad \loopnumber \rightarrow \infty.
\end{align}
Since the number of graphs weighted by symmetry factors is known \cite{borinsky_renormalized_2017,balduf_primitive_2024}, one can instead of the mean equivalently consider the sum of periods, this is the $\loopnumber$-coefficient $\beta^{\text{prim}}_\loopnumber$ of the primitive beta function.  Because of the factorial growth of the number of graphs (and not because of the much slower growth of the numerical value of periods), these coefficients grow factorially with the loop order. To measure their growth rate, one considers a ratio of the form
\begin{align}\label{rL}
	r_\loopnumber :=  {\beta^{\text{prim}}_{\loopnumber+1}} \Big/ \left(\loopnumber \cdot \beta^{\text{prim}}_\loopnumber\right) ~\sim ~1 + \frac{N+10}2 \frac 1 \loopnumber + \mathcal O \left(\loopnumber^{-2}\right) \quad \text{as} \quad \loopnumber \rightarrow \infty.
\end{align}
The precise asymptotics of $r_\loopnumber$ for large $\loopnumber$ is practically relevant because the   perturbation series   $\sum_\loopnumber \beta_\loopnumber^\text{prim}$ is not convergent, and a non-perturbative completion with methods of resummation and resurgence \cite{aniceto_primer_2019,ecalle_guided_2022} needs the large-order behaviour of the perturbative coefficients $\beta^\text{prim}_\loopnumber$ as an input.

\begin{figure}[htb]
	\centering
		\begin{subfigure}{.48\linewidth}
		\begin{tikzpicture}[baseline]
			\begin{axis}[
				width=\linewidth, height=.75\linewidth,
				title={$f_\loopnumber$  as function of $\frac 1 \loopnumber$},
				xlabel={$1/\loopnumber$}, 
				grid=major,
				ymin=1.2, ymax=2.2,
				xmin=0, xmax=0.11,
				xtick={0, .01, .02, .04,0.0555556,0.0769231, .1},
				xticklabels={$\frac 1 \infty$, $\frac 1 {100}$, $\frac 1 {50}$, $\frac{1}{25}$, $\frac 1 {18}$, $\frac 1 {13}$, $\frac 1 {10}$},
				minor y tick num=1,
				yticklabel style={
					anchor=east,
					/pgf/number format/precision=1,
					/pgf/number format/fixed,
					/pgf/number format/fixed zerofill,
				}
				]

				\addplot[darkgreen,samples=10,domain=0:0.95,line width=.9pt] {3/2 + 15/4*x};
			
				\node[darkgreen,rotate=15] at (axis cs:.027,1.67){\footnotesize expected asymptotics};
				\node[black,rotate=30] at (axis cs:.08,2.05){\footnotesize data $N=4$};
				\node[black,rotate=30] at (axis cs:.095,1.95){\footnotesize data $N=1$};
				\node[darkred,rotate=30] at (axis cs:.025,1.45){\footnotesize wrong extrapolation};
				
				\addplot[darkred,dashed,samples=10,domain=0:0.11,line width=.9pt] {1.35*(1 + 5.5*x)};

				\addplot [black, only marks, mark size=1.9pt] table {plotdata/periodRat1.txt};
				\addplot [black, only marks, mark size=1.9pt] table {plotdata/periodRat4.txt};
			\end{axis}
		\end{tikzpicture}
		\subcaption{}
		\label{fig:period_f}
	\end{subfigure}
		\begin{subfigure}{.48\linewidth}
		\begin{tikzpicture}[baseline]
			\begin{axis}[
				width=\linewidth, height=.75\linewidth,
				title={$r_\loopnumber$  as function of $\frac 1 \loopnumber$},
				xlabel={$1/\loopnumber$}, 
				grid=major,
				ymin=.7, ymax=2.2,
				xmin=0, xmax=0.11,
				xtick={0, .01, .02, .04,0.0555556,0.0769231, .1},
				xticklabels={$\frac 1 \infty$, $\frac 1 {100}$, $\frac 1 {50}$, $\frac{1}{25}$, $\frac 1 {18}$, $\frac 1 {13}$, $\frac 1 {10}$},
				minor y tick num=1,
				yticklabel style={
					anchor=east,
					/pgf/number format/precision=1,
					/pgf/number format/fixed,
					/pgf/number format/fixed zerofill,
				}
				]

				\addplot[darkgreen,samples=10,domain=0:0.08,line width=.9pt] {1 + 11/2*x};
				\addplot[darkgreen,samples=10,domain=0:0.08,line width=.9pt] {1 + 7*x};
				
				\node[darkgreen,rotate=18] at (axis cs:.027,1.3){\footnotesize expected asymptotics};
				\node[black,rotate=3] at (axis cs:.08,1.95){\footnotesize data $N=4$};
				\node[black,rotate=28] at (axis cs:.095,1.7){\footnotesize data $N=1$};
				\node[darkred,rotate=27] at (axis cs:.028,.95){\footnotesize wrong extrapolation};
				
				\addplot[darkred,dashed,samples=10,domain=0:0.11,line width=.9pt] {.79*(1 + 16.5*x)};
				\addplot[darkred,dashed,samples=10,domain=0:0.11,line width=.9pt] {.79*(1 + 13.7*x)};

				\addplot [black, only marks, mark size=1.9pt] table {plotdata/betaRat1.txt};
				\addplot [black, only marks, mark size=1.9pt] table {plotdata/betaRat4.txt};
			\end{axis}
		\end{tikzpicture}
		\subcaption{}
		\label{fig:beta_r}
	\end{subfigure}
	\caption{\textbf{(a)} Growth ratio $f_\loopnumber$ (\cref{fL}), plotted as function of $\frac 1 \loopnumber$. \textbf{(b)} Growth ratio $r_\loopnumber$ (\cref{rL}). In both cases,  an extrapolation of the numerical data (red, dashed) does not reproduce the correct  large-$\loopnumber$ asymptotics (green lines). Still, the $N$-dependence of the numerical data has the expected pattern.  }
	\label{fig:growth_rate}
\end{figure}
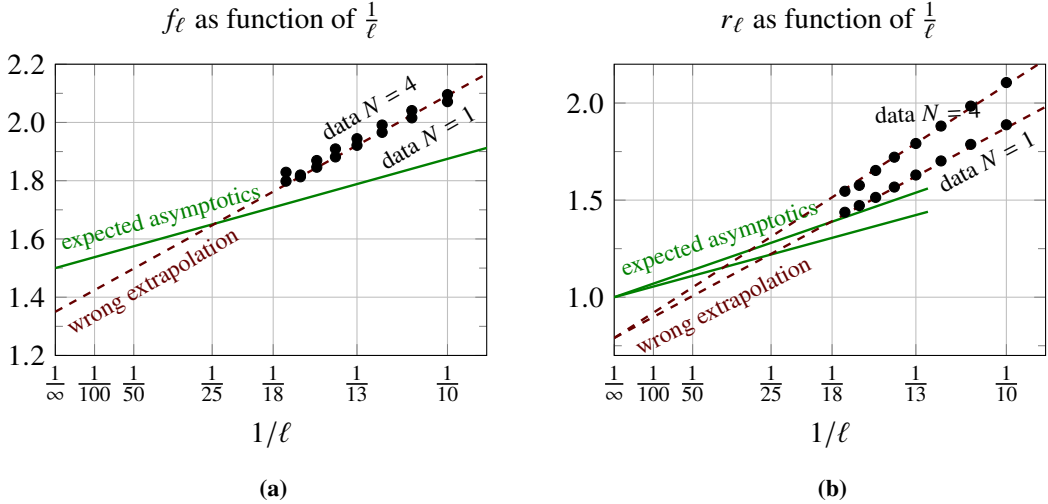

From the plots in \cref{fig:growth_rate}, we observe that the numerical data for $\loopnumber \leq 18$ (black points) does not show the expected asymptotic growth rates (green lines). Interestingly, the pattern of $N$-dependence (i.e. a $\loopnumber \rightarrow \infty$ limit independent of $N$, and a slope linear in $N$ for $r_\loopnumber$, and independent of $N$ for $f_\loopnumber$) still matches the asymptotic expectation.   We know from a comparison to 0-dimensional quantum field theory \cite{balduf_primitive_2024} and from further numerical studies \cite{borinsky_tropicalized_2025} that the asymptotics is indeed accurate; the correct interpretation of \cref{fig:growth_rate} is  that the numerical value of periods approaches its  asymptotic growth rate only upwards of $\loopnumber \approx25$ loops.

\section{Correlations and Predictors}\label{sec:correlation}

\begin{figure}[htbp]
	\begin{subfigure}{.49\linewidth}
		\centering
		\includegraphics[width=.85\linewidth]{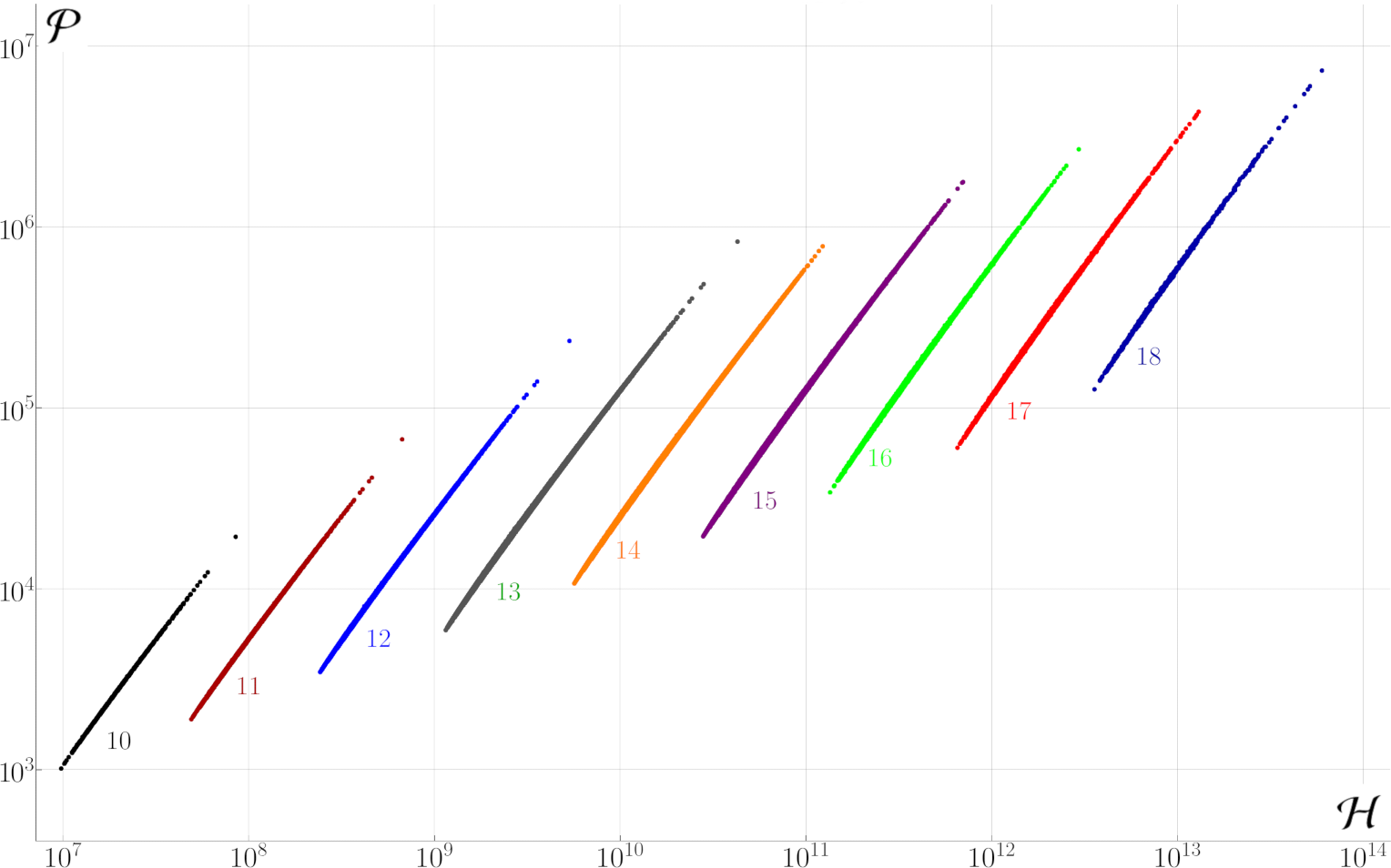}
		\subcaption{}
		\label{fig:correlation_H}
	\end{subfigure}
	\begin{subfigure}{.49\linewidth}
		\centering
		\includegraphics[width=.85\linewidth]{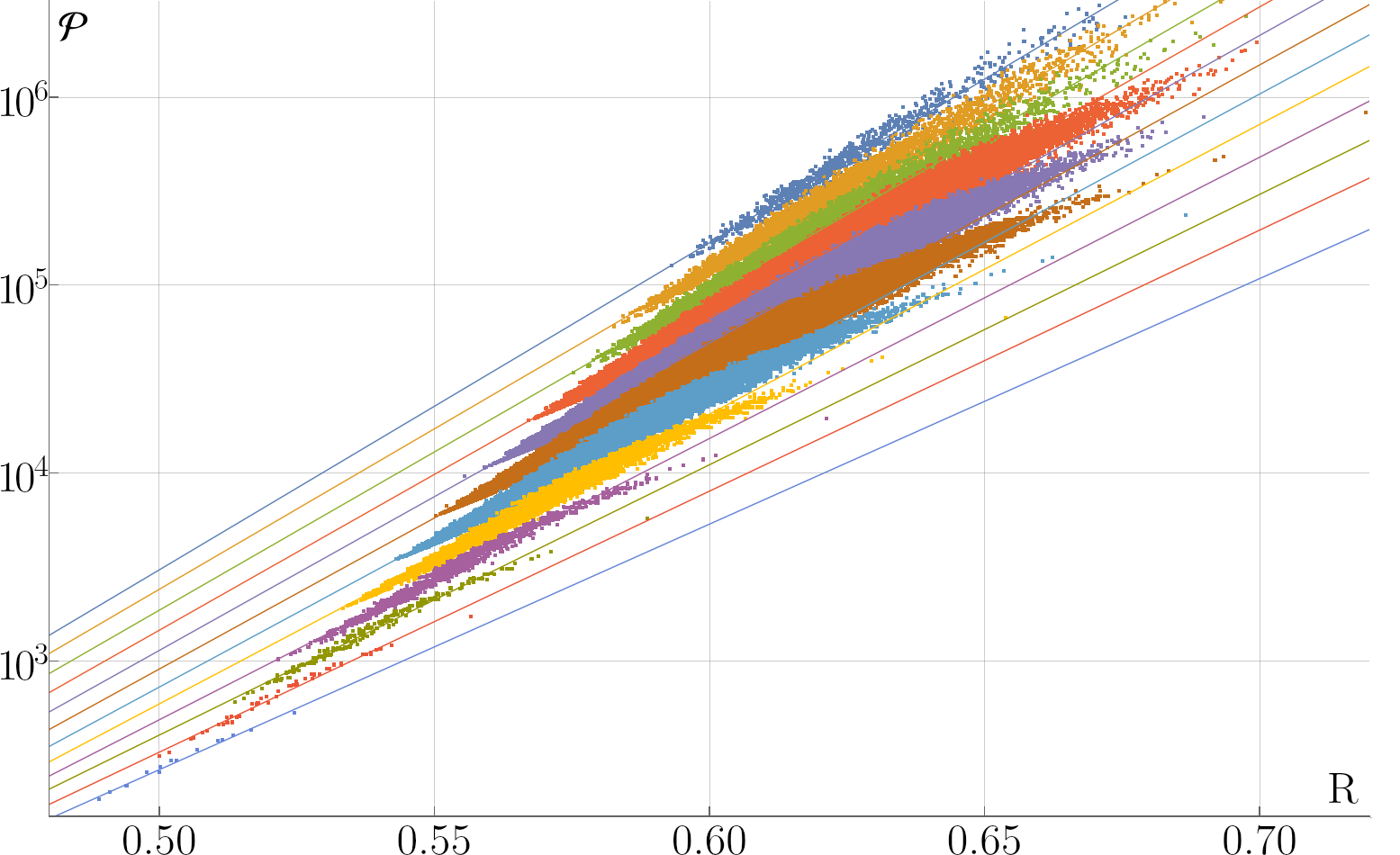}
		\subcaption{}
		\label{fig:correlation_R}
	\end{subfigure}
	\caption{\textbf{(a)} Period as a function of the Hepp bound, for various loop orders. The correlation is very strong. \textbf{(b)} Period as a function of the average resistance. The correlation is weaker, but still clearly visible. }
	\label{fig:correlation}
\end{figure}

Our   data set   shows empirically that the period is correlated with various properties of the underlying graphs. 
The correlation is most striking with the \emph{Hepp bound} $\hepp$ \cite{panzer_hepps_2022} (see \cref{fig:correlation_H}) and the \emph{Martin invariant} \cite{panzer_feynman_2023}, two quantities that were specifically engineered to match the combinatorial properties of Feynman periods, but which are rather expensive to compute for a given graph. In comparison, the counts of certain cycles and cuts in the graph are much faster to compute, but still allow to predict the period to within a few percent accuracy \cite{balduf_predicting_2024}. Another quantity that is more interesting from a physics perspective is the \emph{Kirchhoff index}, or mean electrical resistance. Pictorially, one replaces every edge of a completion by a unit electrical resistor, then the Kirchhoff index is the mean of the effective resistances between any pair of vertices; it can be computed in less than one millisecond for a 20-loop graph.  The correlation between $\period$ and $R$ (see \cref{fig:correlation_R}) is weaker than for $\period$ and $\hepp$, but it still determines the period with an average error less than 10\%.

These correlations have important practical applications for  importance sampling of  Feynman graphs prior to integration. In \cite{balduf_predicting_2024}, we implemented such a sampling algorithm and found empirically that at $\loopnumber =14$ it is roughly 1000-times faster to compute the sum of all integrals, compared to uniform random sampling. This algorithm has also been used to compute the $N$-dependent data for \cite{balduf_primitive_2024}.

\section{Outlook}

At large loop order, the numbers of Feynman graphs quickly range in the billions, which makes it impossible to compute every individual integral. With a relatively naive importance sampling algorithm, we have computed the sum of periods to 4-digit numerical accuracy while only evaluating a tiny fraction $\ll 1\%$ of the integrals,  speeding up the computation by 3 orders of magnitude. An even larger performance gain would be achievable by using a more refined prediction function. 

Beyond just periods,  there is no  reason to believe that similar correlations would be absent in more general Feynman integrals. Given the steady progress in   pattern recognition and machine learning, it appears very possible that entire Feynman amplitudes could be computed by clever statistical sampling in a data-driven way. If anything, such methods tend to perform better the larger the data set is, thereby, we believe that the numerical computation of Feynman amplitudes at large loop order is not as unsurmountable a challenge as it might as first appear.

\bibliographystyle{JHEP}
\bibliography{Talk_2025_Statistics}

\end{document}